# Modelling Evaluation of MPLS using Physical and virtual Network on GNS3

1Abdul Ahad Abro 2Abdul Basit Abro 3Mehvish Abro 4Asad Aslam Siddique

**Abstract:** The Multi-Protocol Label Switching (MPLS) is an emerging technology which have quality, effectiveness and administration quality. On the contrary, traditional network inside part passage steering conventions ruin the viable acknowledgment of modern activity designing approaches in legacy IP systems. Virtualization of the network could easily be assuring the network performance and virtual network are logically connected with one physical machine so that data could easily be send and get information from one virtual machine to the next machine. The purpose of this Paper is to analyse the traffic of MPLS using physical and virtual networks. This Paper will show that MPLS could also be run on physical and virtual networks. MPLS is running nowadays to provide local area network speed into the wide area Network.

## 1. INTRODUCTION

The motivation behind this paper is to dissecting the MPLS network. MPLS network is divided into two parts; Core network is used to connect one side of the customer end to another side. Cloud in the diagram shows that the core network is depended on the virtual routers that connected in a mesh. [7] Using MPLS, we could increase the efficiency of the packet passing through them [7]. This figure1 shows that MPLS can work on both virtual & emulated network, while both devices can work in a same network by connecting virtually with one another.

Access network is a network when traffic analyzing starts from one side of the customer End too other side. We will analyses traffic performance on both virtual & physical network to verify, which one would be best to test.

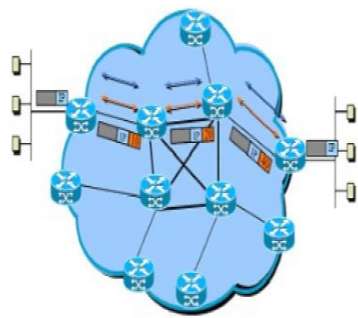

Figure1: Working of MPLS

*Corresponding Author email: abrobasit@hotmail.com    mehvishabro03@gmail.com

Traffic engineering, at its central part, is the art of moving traffic around so traffic from a jam-packed link is moved onto the unused capacity on a different link. Traffic engineering is by no means an MPLS-specific thing; it's a universal observation. [2] Traffic engineering can be put into practice by something as simple as alteration IP metrics on interfaces, or something as multifarious as running an ATM PVC full-mesh and re- optimizing PVC paths based on traffic needs across [2]. MPLS with TE is an effort to obtain the finest connection-oriented traffic engineering techniques (such as ATM PVC placement) and unite them with IP routing. The theory telling here is that doing traffic engineering with MPLS can be as effective as with ATM, but without many shortcomings of IP over ATM.

## 2. Virtual Networking

Virtual networking is defined as the network setting in the virtual box. Virtual box supports eight virtual Ethernet cards for each virtual machine and it can be setup individually or as well as the hardware that will be virtualized. Virtual cards operate according to the physical hardware on the host in the virtualization mode. Virtual box provides four network interface card and can be configured through Graphical user interface. There are various networks setting in the network Modes which provides by Virtual box. Network Address translation is a way to communicate external network from virtual machine.it doesn't required any configuration in the network on guest and host OS.

In bridged Networking, Virtual box does use the tool drivers of device to filter the information or data on host system from physical network adapter. It's also referred to as net filter tool drivers.

For internal networking, no predefine configuration because it's created automatically in the network. There is no difference in internal networking and bridging networking only is security benefit for using internal networking. Internal networking and bridging networking, all VM can exchange information and communication with the external network. Host-only networking is the additional networking mode which added in the Virtual Box. [9] It is often thought of as a central piece between the bridged and internal networking modes. [9] In bridged networking, the virtual machines will consider to one another and also the host as if they are connected with the physical LAN switch. [5] Similarly, like internal networking but, Virtual machine must connect to the physical interface to communicate with the outside network [5].

Generic networking is UDP tunnel networking mode permits interconnecting virtual machines running on different hosts. This networking can be done by encapsulating the frame into UDP/IP datagrams or packet which send and receive by the guest network card and send them on any available network.

*Corresponding Author email: abrobasit@hotmail.com   mehvishabro03@gmail.com

Virtual Distributed Ethernet (VDE3) supports various flavors in virtual network infrastructure system, providing across multiple hosts in a secure manner. Its perform layer2 and layer3 switching, including STP, VLAN, VLAN tagging and WAN switching.

## 3. Related Work

Utilizing MPLS exchanging method the correspondence might be improved by labeling of marks between OSI layers of Data-link and Network that are supporting a few characteristics in Traffic Engineering. This Paper and virtual network because of increasing demand of virtualization to reduce power and cost and ensure the reliability. The Traffic analysis and path protection could be shown in simulated environment while sending traffic from client end.

## 4. Simulation

Testing and performance of Physical and virtual network by using GNS3. Mikrotik. Open software solutions and cisco platform performing the same scenario on GNS3. Mikrotik. Virtual Operating system are the leverage standard x.86 hardware can be viable alternatives to expensive proprietary solutions. Mikrotik open solution demonstrates that standard hardware at attractive price points can be one of the good platforms than a purpose-built box from a leading vendor, but it also provides enough processing headroom for expansion. [3] The structural economics of the open model provides users with a cost-effective way to scale performance. Another important benefit of Mikrotik open solution is that it decouples software from the underlying hardware and allows users to achieve software feature and service extensibility by leveraging thousands of Linux-compatible application packages that can integrate with the Mikrotik software [3].

I have used an x.86 machine (desktop/laptop) QUAD core to run Mikrotik virtual OS router by using GNS3 virtual box and all vm are connected with one another. On the other hand, I have used a Cisco router (GNS3) which is directly connected to the other cisco router's Ethernet interface to exchange the routing information. The following are the hardware and software configurations on both the Mikrotik and Cisco routers:

### Scenario 1 Physical Platform

*Corresponding Author email: abrobasit@hotmail.com   mehvishabro03@gmail.com

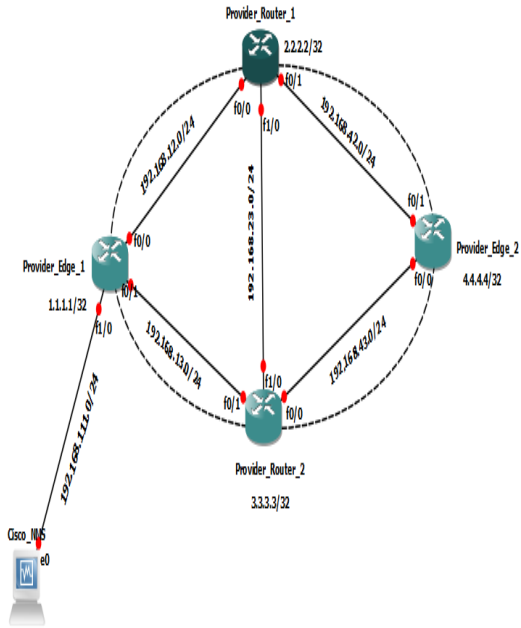

Figure2: showing the test topology used for the testing of Cisco and its performance graph.

## Scenario 2: (Mikrotik Linux virtual platform)

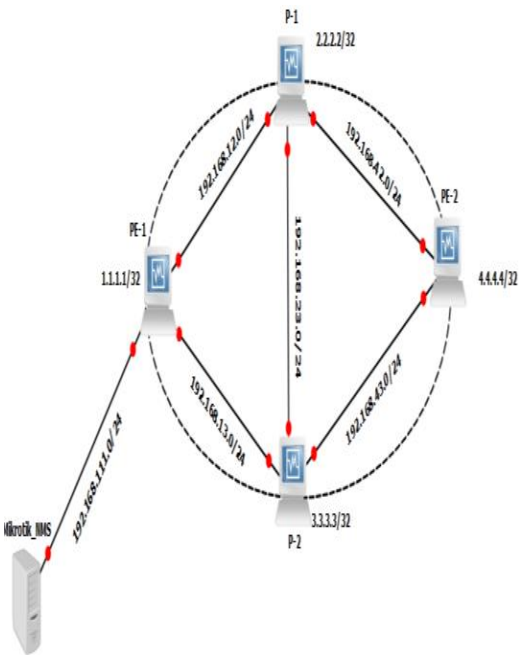

Figure3: showing the test topology used for the testing of Mikrotik virtual OS and its performance

## 5. Performance Evaluation

Performance graphs captured on the monitoring server as a result of sending packets from the source to the respective destinations.

## 6. Throughput

In physical network, data sent about 1Mbps and its start from source that is traffic in which is fast Ethernet 0/1, peak traffic in is 752kbps but its average is about 522kbps while traffic out which is fast Ethernet 0/0, peak throughput is 764kbps but its average is about 530kbps.

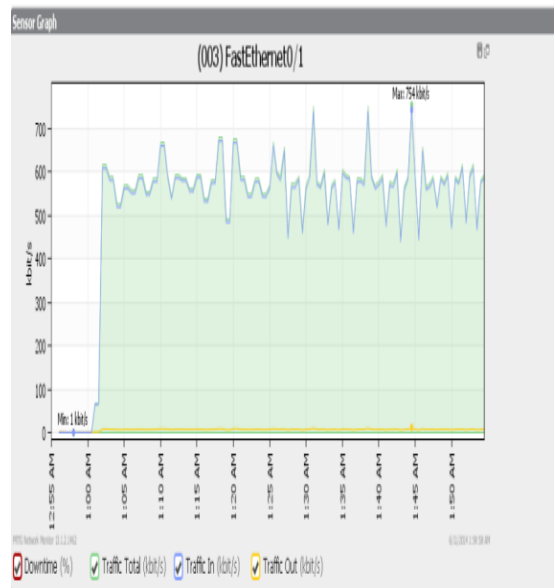

*Corresponding Author email: abrobasit@hotmail.com   mehvishabro03@gmail.com

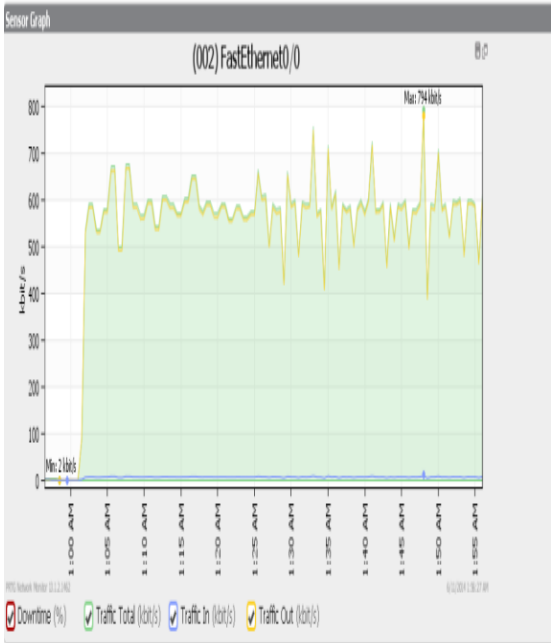

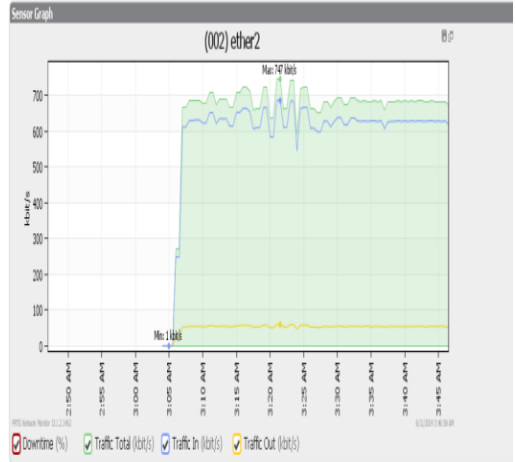

Figure4: Throughput of physical network

In virtual network, data sent about 1Mbps and its start from source that is traffic in which is ether 2, peak traffic in is 748kbps but its average is about 642kbps while traffic out which is Ethernet 2, peak throughput is 1021kbps but its average is about 575kbps.

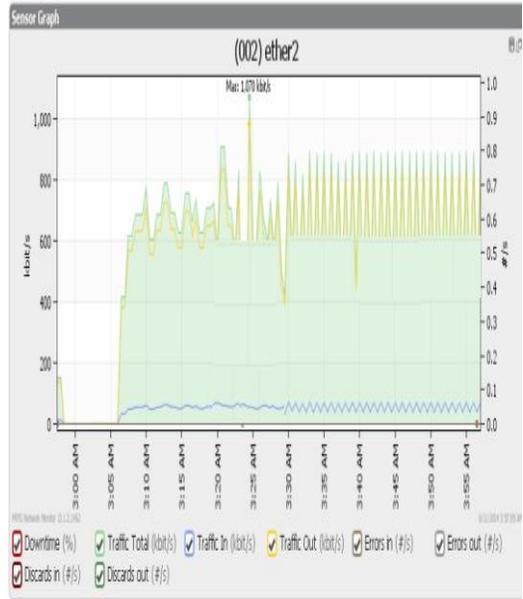

Figure5: Throughput of virtual network

*Corresponding Author email: abrobasit@hotmail.com  mehvishabro03@gmail.com

## 7. Delay

In physical network, data sent about 1Mbps and we found so much delay in that network because its reached at maximum level that is 121msec and its minimum delay level is 30msec.

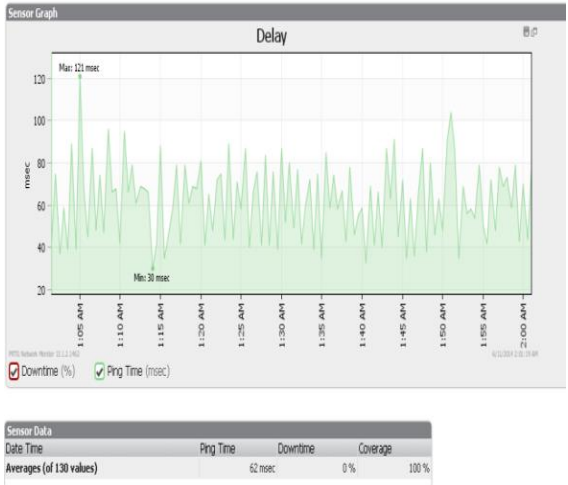

Figure6: delay of physical network

In virtual network, data sent about 1Mbps and we found so much delay in that network because its reached at maximum level that is 8msec and its minimum delay level is 3msec.

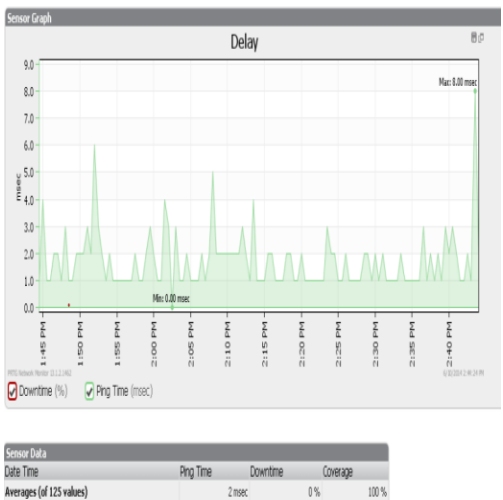

Figure7: Delay of virtual network

## 8. CPU utilization

In Physical network, CPU utilization was directly shot up to 12% then it had been stable at 16%. It directly put impact on the hardware performance.

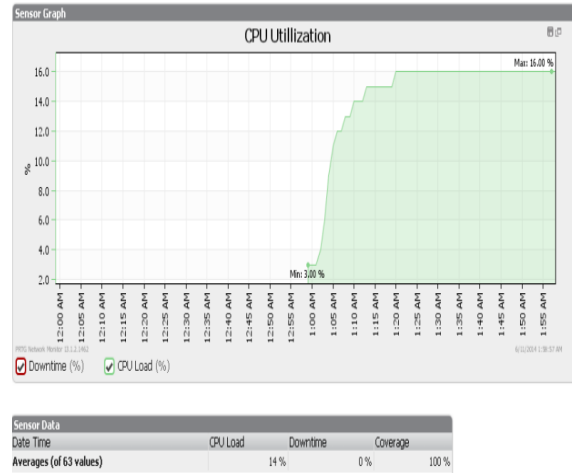

Figure8: CPU utilization of physical network

In virtual network, CPU utilization was increasing slowly from 2% then it had been stable at 5%. It directly put impact on the hardware performance.

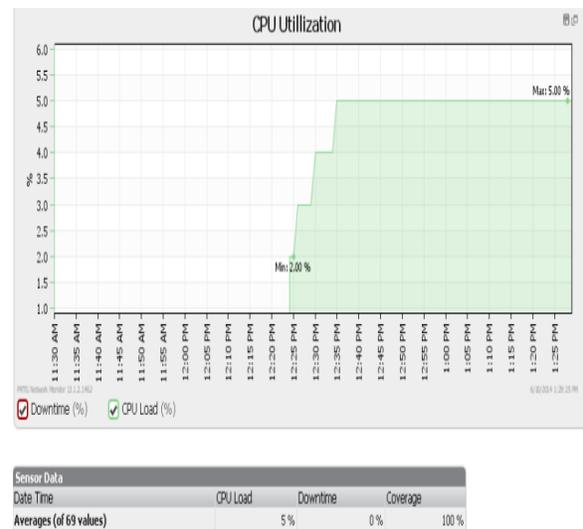

Figure9: CPU utilization of virtual network

*Corresponding Author email: abrobasit@hotmail.com   mehvishabro03@gmail.com

## 9. Convergence Time

In physical network, convergence time taken place within a second but only found the difference in seconds that's 31second.

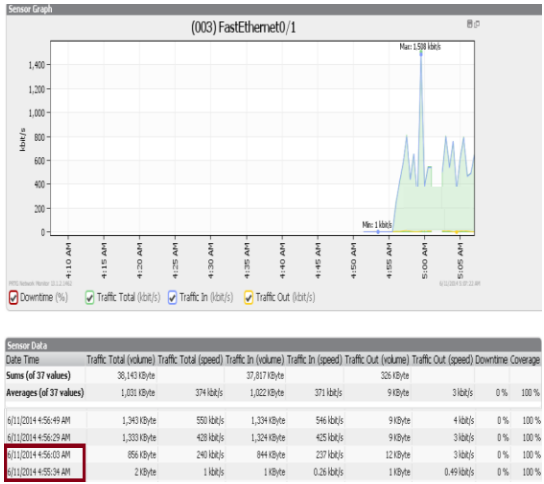

Figure10: Convergence time of physical network

In virtual network, convergence time taken place within a second but only found the difference in seconds that's 26second.

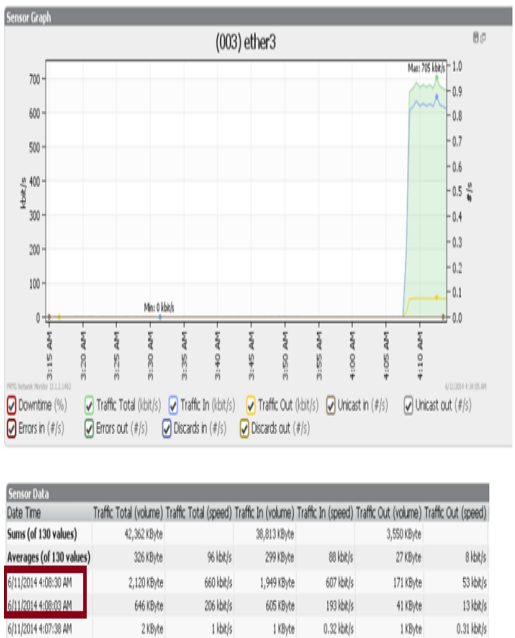

Figure11: Convergence time of virtual network

## 10. Conclusion

The purpose of this paper was to investigate the better performance of MPLS either on physical or virtual network. The tested mechanism is that virtual & physical network can easily be integrated and MPLS can be implemented on both virtual & physical infrastructures of the network. The result shows of both physical and virtual network, bandwidth and delay could easily be optimized and have the same effect as the real environment. Virtual router helps to reduce physical hardware cost as well as the best resource utilization and also less power consumed in the network.

## 11. Future Work

Software Defined Networking (SDN) is a rising building design that is sensible, financially better, and versatile, Showing it perfect for the high-data transmission, and multicolored nature for the current requirement of the market application. This structural engineering mixed the Network control and sending capacities empowering the Network control to end up specifically programmable and the below foundation to be disconnected for provisions and system administrations. The Open flow convention protocol is a base component for building SDN architecture.

*Corresponding Author email: abrobasit@hotmail.com   mehvishabro03@gmail.com

*Corresponding Author email: abrobasit@hotmail.com    mehvishabro03@gmail.com